\documentclass{article}
\usepackage{spconf,amsmath,graphicx,enumitem,IEEEtrantools,pifont,booktabs}
\usepackage{soul,tablefootnote, verbatim,pifont,amsfonts,multirow}

\title{Diffusion-Based Speech Enhancement with \\ Joint Generative and Predictive Decoders}
\name{Hao Shi$^{\dagger,*}$\thanks{*Work done during an internship at Sony Group Corporation},~Kazuki Shimada$^{\ddagger}$,~Masato Hirano$^{\S}$,~Takashi Shibuya$^{\ddagger}$,~Yuichiro Koyama$^\S$,\\ \em Zhi Zhong$^\S$,~Shusuke Takahashi$^\S$,~Tatsuya Kawahara$^\dagger$,~Yuki Mitsufuji$^{\ddagger,\S}$}

\address{
 $^\dagger$Graduate School of Informatics, Kyoto University, Kyoto, Japan\\
 $^\ddagger$Sony AI, Tokyo, Japan $^\S$Sony Group Corporation, Tokyo, Japan
}
\begin{document}
\ninept
\maketitle

\begin{abstract} 
Diffusion-based generative speech enhancement (SE) has recently received attention, but reverse diffusion remains time-consuming. 
One solution is to initialize the reverse diffusion process with enhanced features estimated by a predictive SE system. 
However, the pipeline structure currently does not consider for a combined use of generative and predictive decoders. 
The predictive decoder allows us to use the further complementarity between predictive and diffusion-based generative SE. 
In this paper, we propose a unified system that use jointly generative and predictive decoders across two levels. 
The encoder encodes both generative and predictive information at the shared encoding level. 
At the decoded feature level, we fuse the two decoded features by generative and predictive decoders. 
Specifically, the two SE modules are fused in the initial and final diffusion steps: the initial fusion initializes the diffusion process with the predictive SE to improve convergence, and the final fusion combines the two complementary SE outputs to enhance SE performance. 
Experiments conducted on the Voice-Bank dataset demonstrate that incorporating predictive information leads to faster decoding and higher PESQ scores compared with other score-based diffusion SE (StoRM and SGMSE+). 
\end{abstract}

\vspace{-5pt}
\section{Introduction}
\vspace{-5pt}
Speech enhancement (SE) aims to recover clean speech from noisy signals. 
Since noise in real-world scenarios \cite{6732927,saito2021training,uhlich2020open} significantly degrades the performance of speech applications, SE is an important front-end in speech processing applications, such as automatic speech recognition \cite{45168, 6639038, 6843279}, speaker identification \cite{BAI202165}, and semantic communication \cite{9679803,grassucci2023generative,jiang2023large,grassucci2023enhancing}. 
Supervised SE systems \cite{9103053, 6932438} have more robust performance compared to traditional SE systems \cite{5745591,397090}. 
Therefore, they have recently been intensively investigated \cite{Yin_Luo_Xiong_Zeng_2020, 8966946}. 
They can be classified into two types: predictive (also called discriminative) \cite{9054661,https://doi.org/10.48550/arxiv.2104.03538,8707065,LI2022108499} and generative \cite{Pascual2017, 9746901, https://doi.org/10.48550/arxiv.2208.05830, https://doi.org/10.48550/arxiv.2212.11851}. 
They adopt different paradigms. 
Predictive SE systems learn the single best deterministic mapping between noisy speech and its corresponding clean speech \cite{https://doi.org/10.48550/arxiv.2212.11851}. 
In contrast, the target distribution of clean speech is implicitly or explicitly learned in generative SE systems \cite{https://doi.org/10.48550/arxiv.2212.11851}. 
Generative SE models include variational auto-encoders \cite{8516711}, generative adversarial networks \cite{Pascual2017} and diffusion models \cite{9746901, https://doi.org/10.48550/arxiv.2208.05830, https://doi.org/10.48550/arxiv.2212.11851}. 
Among them, diffusion models have recently attracted a significant attention due to their success in other fields \cite{NEURIPS2019_3001ef25, NEURIPS2021_b578f2a5, 10095046}.

Diffusion models are inspired by non-equilibrium thermodynamics. 
The data are gradually turned into noise, and the neural network learns to invert the progressive noise-adding process. 
The conditional diffusion model \cite{9746901, hu2023noiseaware} uses noisy spectrograms as the conditioner. 
However, its objective function assumes that the global distribution of the additive noise follows a standard white Gaussian distribution, which is inconsistent with real noise statistics. 
The score-based diffusion model \cite{https://doi.org/10.48550/arxiv.2208.05830} is based on stochastic differential equations (SDEs), which makes the training fully generative without any prior noise distribution assumptions. 
The reverse diffusion process (decoding) of the diffusion models is very time-consuming. 
To reduce the number of reverse diffusion steps, several studies \cite{https://doi.org/10.48550/arxiv.2212.11851,sawata23_interspeech} have combined a predictive model with the generative model. 
A previous work use the predictive information as the initialization for the generative model \cite{https://doi.org/10.48550/arxiv.2212.11851}. 
The work also show that the generative and predictive models have different distortions \cite{https://doi.org/10.48550/arxiv.2212.11851}.
Furthermore, UNIVERSE \cite{universal_2022} shows that adding predictive information to the decoder of the diffusion model can help diffusion score estimation. 
However, the previous pipeline structures \cite{https://doi.org/10.48550/arxiv.2212.11851,sawata23_interspeech} limit the systems further use the complementarity between the generative and predictive modules.

In this paper, we propose a unified speech enhancement (SE) system that integrates generative and predictive SE modules at the shared encoding and enhanced feature levels. 
At the shared encoding level, the model incorporates a shared encoder along with both predictive and generative decoders. 
The generative module is a score-based diffusion model adopted \cite{song2021scorebased}. 
To leverage the complementarity between the two modules at the enhanced feature level, we fuse the enhanced generative-based and predictive-based features during the first and final diffusion steps. 
The first step fusion utilizes the predicted enhanced feature to initialize the subsequent diffusion processes. 
In order to maintain small changes in the feature distribution, the two features are fused instead of using the predicted spectra directly, although the enhanced predictive feature has higher performance in the first step than the enhanced generative feature. 
Since the two systems introduce different signal distortions, feature fusion is also adopted in the final step to leverage the complementarity between the generative and predictive SE modules.

\section{Score-based Diffusion Model}
\vspace{-5pt}
\subsection{Stochastic process}
\vspace{-5pt}
The linear stochastic differential equation (SDE) relies on a stochastic diffusion process $\left \{ x_{t} \right \}_{t=0}^{T} $ \cite{song2021scorebased}: 
\begin{equation}
	\setlength{\abovedisplayskip}{0pt}
	\setlength{\belowdisplayskip}{0pt}
	\mathrm{d}x_{t} = \underset{:=\mathrm{f}(x_{t}, y) }{\underbrace{\gamma(y-x_{t})\mathrm{d}t }} +  \underset{:=g(t)}{\underbrace{\left [  \sigma_{min}(\frac{\sigma_{max}}{\sigma_{min}} )^{t} \sqrt{2\mathrm{log} (\frac{\sigma_{max}}{\sigma_{min}} )}  \right ]}}\mathrm{d}\mathrm{w}
	\label{eq1}
\end{equation}
where $x_{t}$ is the current state of the process indexed by a continuous time variable $ t \in  \left [ 0, T \right ]  $  \cite{song2021scorebased}.
$x_{0}$ is the clean speech, which represents the initial condition, 
$y$ is the noisy speech, and $\mathrm{w}$ denotes a standard Wiener process.
The vector-valued function $\mathrm{f}(x_{t}, y)$ is referred to as the drift coefficient, 
$g(t)$ is the diffusion coefficient of $x_{t}$, 
and $\gamma$ is the stiffness. 
$\sigma_{min}$ and $\sigma_{max}$ control the amount of Gaussian white noise at each diffusion timestep. 
The SDE in (\ref{eq1}) has an associated reverse SDE \cite{song2021scorebased,ANDERSON1982313}, 
\begin{equation}
	\setlength{\abovedisplayskip}{0pt}
	\setlength{\belowdisplayskip}{0pt}
	\mathrm{d}x_{t} = \left [ -\mathrm{f}(x_{t},y) + g(t)^{2} \bigtriangledown_{x_{t}}\mathrm{log}p_{t}(x_{t}) \right ] \mathrm{d}t + g(t)\mathrm{d}\bar{\mathrm{w}}   
	\label{eq2}
\end{equation}
Practically, the score $ \bigtriangledown_{x_{t}}\mathrm{log}p_{t}(x_{t})$ is estimated by a score model. 
The score model can be denoted as $s_{\theta } (x_{t}, y, t) $, which is parameterized by a set of DNN parameters $\theta$. 
It receives the current state of the process $x_{t}$, the noisy speech $y$, and the current timestep $t$ as inputs. 
Finally, by substituting the score model into the reverse SDE in (\ref{eq2}) \cite{NEURIPS2021_c11abfd2}, we obtain 
\begin{equation}
	\setlength{\abovedisplayskip}{0pt}
	\setlength{\belowdisplayskip}{0pt}
	\mathrm{d}x_{t} = \left [ -\mathrm{f}(x_{t}, y) + g(t)^{2}s_{\theta } (x_{t}, y, t) \right ] \mathrm{d}t + g(t)\mathrm{d}\bar{\mathrm{w} }   ]
	\label{eq3}
\end{equation}
which can be solved with various solver procedures.

\vspace{-5pt}

\subsection{Training objective}
\vspace{-5pt}
The mean and variance of the process state $x_{t}$ can be derived when its initial conditions are known \cite{oksendal2003}. 
Since the feature used in this paper is a complex spectrogram, at an arbitrary timestep $t$, $x_{t}$ can be directly sampled by $x_{0}$ and $y$ with the perturbation kernel: 
\begin{equation}
	\setlength{\abovedisplayskip}{0pt}
	\setlength{\belowdisplayskip}{0pt}
	p_{0t}(x_{t}|x_{0}, y) = \mathcal{CN}_{\mathbb{c}}(x_{t}; \mu(x_{0}, y, t), \sigma(t)^2 \mathrm{I})
	\label{eq4}
\end{equation}
where $\mathcal{CN}_{\mathbb{c}}$ denotes the circularly symmetric complex normal distribution. 
$\mathrm{I}$ is the identiry matrix. 
The mean and variance can be estimated as follows \cite{oksendal2003}: 
\begin{equation}
	\setlength{\abovedisplayskip}{0pt}
	\setlength{\belowdisplayskip}{0pt}
	\mu \left ( x_{0}, y, t \right ) = e^{-\gamma t} x_{0} + (1 - e^{-\gamma t})y
\end{equation}

\begin{equation}
	\setlength{\abovedisplayskip}{0pt}
	\setlength{\belowdisplayskip}{0pt}
	\sigma (t)^2 = \frac{\sigma^2_{min}((\sigma_{max}/\sigma_{min})^{2t} - e^{-2\gamma t})\mathrm{log}(\sigma_{max}/\sigma_{min})  }{\gamma + \mathrm{log}(\sigma_{max}/\sigma_{min}) } 
\end{equation}
The denoising score matching is described as follows \cite{https://doi.org/10.48550/arxiv.2208.05830}: 
\begin{equation}
	\setlength{\abovedisplayskip}{0pt}
	\setlength{\belowdisplayskip}{0pt}
	\begin{aligned}
		\bigtriangledown_{x_{t}} &\mathrm{log}p_{0t}(x_{t}|x_{0},y) =  \bigtriangledown_{x_{t}}\mathrm{log}\left [    \left | 2\pi \sigma \mathrm{I} \right | ^{-\frac{1}{2}} e^{-\frac{ || x_{t}-\mu ||^{2}_{2}  }{2\sigma ^2}} \right ] \\
		& = \bigtriangledown_{x_{t}}\mathrm{log}\left | 2\pi \sigma (t)\mathrm{I}  \right |^{-\frac{1}{2} } - \bigtriangledown_{x_{t}}\frac{|| x_{t} - \mu (x_{0}, y, t) ||^{2}_{2} }{2\sigma (t)^2} \\
		& = - \frac{x_{t} - \mu (x_{0}, y, t)}{\sigma (t)^2}  
	\end{aligned}
	\label{eq9}
\end{equation}
At each training step, the following four steps are executed \cite{https://doi.org/10.48550/arxiv.2208.05830}: 
\begin{enumerate}[topsep=0pt, fullwidth, itemindent=1.5em]
	\item [\ding{172}] Sample a random $t \sim   \mathcal{U}\left [ t_{\varepsilon}, T \right ]  $;
	\item [\ding{173}] Sample $(x_{0}, y)$ from the dataset;
	\item [\ding{174}] Sample $\mathrm{z} \sim  \mathcal{CN}(\mathrm{z}; 0, \mathrm{I} ) $;
	\item [\ding{175}] Sample $x_{t}$ from (\ref{eq4}) by computing:
	\begin{equation}
		\setlength{\abovedisplayskip}{0pt}
		\setlength{\belowdisplayskip}{0pt}
		x_{t} = \mu (x_{0}, y, t) + \sigma (t) \mathrm{z} 
		\label{eq10}
	\end{equation}
\end{enumerate} 
The training loss is computed between the model output and the score of the perturbation kernel. 
By substituting (\ref{eq10}) into (\ref{eq9}), the overall training objective is described as follows \cite{https://doi.org/10.48550/arxiv.2208.05830}:
\begin{equation}
	\setlength{\abovedisplayskip}{0pt}
	\setlength{\belowdisplayskip}{0pt}
	\underset{\theta }{\mathrm{arg}~\mathrm{min}\,} \mathbb{E}_{t, (x_{0}, y), \mathrm{z}, x_t|(x_0,y) }\left [ ||  \mathrm{s}_\theta(x_t,y,t) +  \frac{\mathrm{z} }{\sigma (t)}  ||^2_2 \right ]  
	\label{genloss}
\end{equation}

\vspace{-5pt}

\begin{figure}[htpb]
	\centering
	\includegraphics[width=0.4\textwidth]{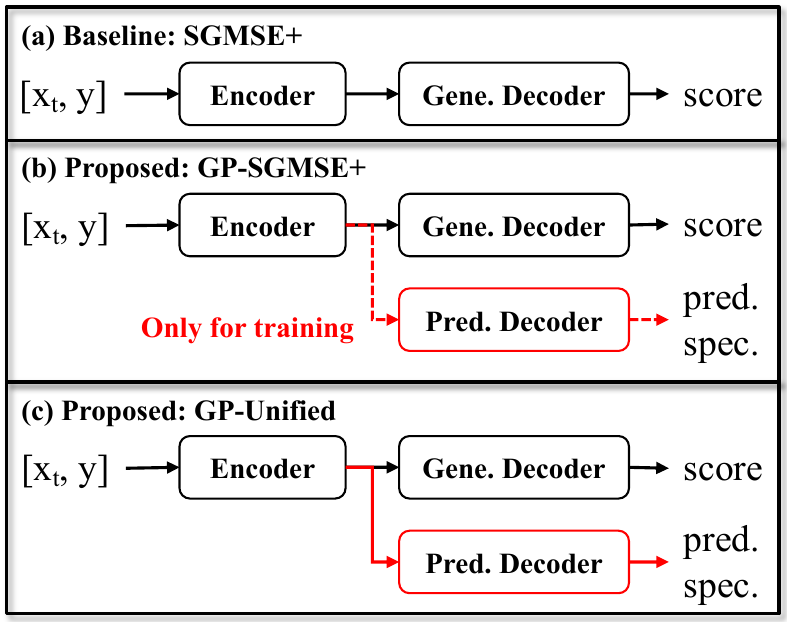}
	\vspace{-10pt}
	\caption{The score model (used in PC samplers) structure: \\(a) the baseline SGMSE+; (b) the proposed Generative and Predictive based SGMSE+ (GP-SGMSE+); (c) the proposed Unified Generative and Predictive model (GP-Unified). \textbf{Note that the skip connection exists between the encoder and decoders (generative and predictive).}}
	\vspace{-15pt}
	\label{flowchart}
\end{figure}

\vspace{-5pt}

\subsection{Inference}
\vspace{-5pt}
For inference, a trained score model $\mathrm{s}_\theta  $ approximates the true score for all $t\in \left [ 0, T \right ] $. 
The noisy speech $y$ is conditioned to estimate clean speech $x_0$ by solving the plug-in reverse SDE in (\ref{eq3}). 
The initial condition of the reverse process at $t = T$ can be determined as follows \cite{https://doi.org/10.48550/arxiv.2208.05830}:
\begin{equation}
	\setlength{\abovedisplayskip}{0pt}
	\setlength{\belowdisplayskip}{0pt}
	x_T \sim \mathcal{N}_\mathbb{C} (x_T; y, \sigma (T)^2\mathrm{I} )
\end{equation}
The denoising process through the reverse process starts at $t=T$ and ends at $t = 0$ iteratively. 
PC samplers combine single-step methods with numerical optimization approaches for the reverse SDE \cite{song2021scorebased}. 
PC samplers consist of a predictor and a corrector. 
The predictor solves the reverse process by iterating through the reverse SDE \cite{song2021scorebased}. 
The corrector refines the current state after each iteration step of the predictor \cite{song2021scorebased}. 

\vspace{-10pt}

\section{Proposed Method}
\begin{figure}[htpb]
	\centering
	\includegraphics[width=0.47\textwidth]{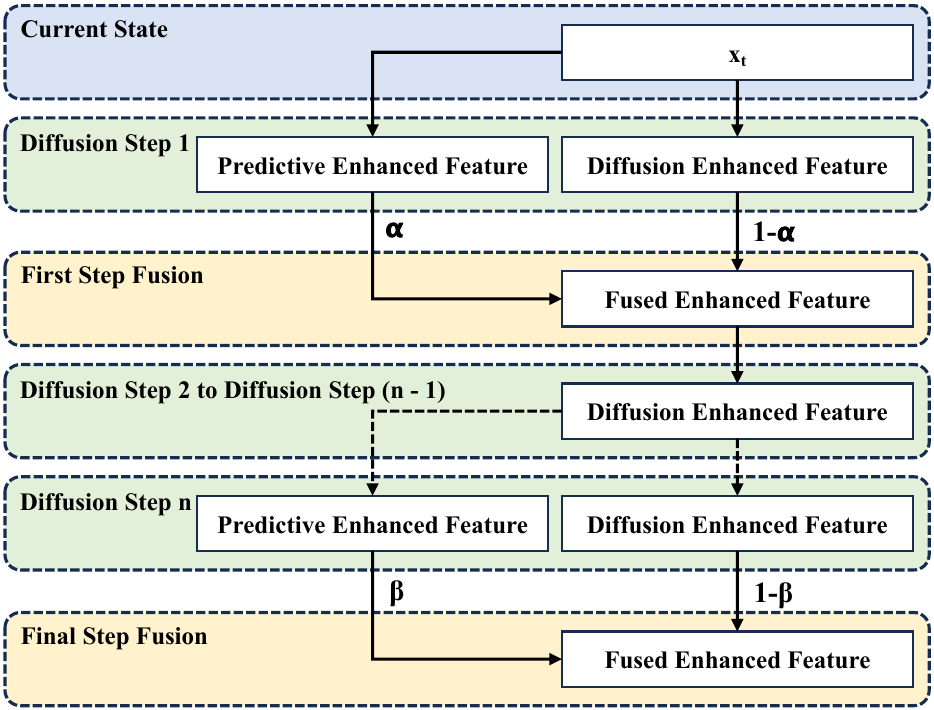}
	\vspace{-10pt}
	\caption{A flowchart of the proposed method. Generative and predictive SE systems are fused in the first and final diffusion steps.}
	\vspace{-15pt}
	\label{pipline}
\end{figure}

\vspace{-5pt}
Score-based diffusion models have already achieved performance comparable to that of the predictive model. 
However, the predictive model calls the neural network only once, while the diffusion model needs to call the neural network several times, which significantly increases the decoding time. 
Some multi-stage models incorporate enhanced predictive features into the diffusion model to significantly reduce the number of diffusion steps. 
However, these systems are pipeline structures, which limit to further utilize the complementarity between the predictive and generative models, since the generative models distort signals much differently from how the predictive models do \cite{https://doi.org/10.48550/arxiv.2212.11851}. 
Besides, introducing predictive information into generative models can help improve the performance of the diffusion process \cite{universal_2022}. 
Therefore, we implement these two different SE systems in the unified system by fusing them. 
The flowchart of the proposed method is shown in Fig.~\ref{pipline}.

\vspace{-5pt}

\subsection{GP-SGMSE+}
\vspace{-5pt}
The structure of the generative- and predictive-based model (\textbf{GP-SGMSE+}) to estimate a score $ \bigtriangledown_{x_{t}}\mathrm{log}p_{t}(x_{t})$ is shown in Fig.~\ref{flowchart}(b). 
The model contains a shared encoder and two decoders. 
The original SGMSE+ neural network \cite{https://doi.org/10.48550/arxiv.2208.05830} contains an encoder and a decoder. 
And the encoder only focuses on encoding generative information. 
In GP-SGMSE+, the generative and predictive decoders share an encoder. 
In this model, the predictive decoder is only used during training to introduce predictive enhancement information into the model. 
When reverse diffusion process, no additional parameters are introduced to the baseline model shown in Fig.~\ref{flowchart}(a). 
The mean square error is used for computing the predictive loss:
\begin{equation}
	\setlength{\abovedisplayskip}{0pt}
	\setlength{\belowdisplayskip}{0pt}
	L_{pred} = ||  x_{pred}  - x_0   ||^2
	\label{preloss}
\end{equation}
where $x_{pred} $ is the output of the predictive decoder. 
The final loss combines both the predictive and generative loss in (\ref{genloss}) and (\ref{preloss}). 
Because the two tasks are equally important, the weights of the losses of the two parts are 0.5 during training. 
The weights of the losses will not affect the two separate decoders, but will affect the shared encoder information, which will be explored in the future. 
During reverse diffusion process, only the generative decoder is used. 
The inference process is the same as in Section~2.3.

\subsection{GP-Unified}
\vspace{-5pt}
The structure of the unified generative and predictive model (\textbf{GP-Unified}) is shown in Fig.~\ref{flowchart}(c). 
Unlike in ``GP-SGMSE+'', the predictive decoder is also used in the reverse diffusion process. 
During the reverse diffusion process, the enhanced generative and predictive features are fused in the first and final diffusion steps. 
The first step fusion is to use the predictive enhanced spectrogram to initialize the follow-up processes of diffusion: 
\begin{equation}
	\setlength{\abovedisplayskip}{0pt}
	\setlength{\belowdisplayskip}{0pt}
	\widehat{x_{1}} = \alpha * x_{1} +  (1 - \alpha) * x_{1}^{pre}
\end{equation}
where $\widehat{x_{1}} $ is the first diffusion-enhanced complex spectrogram, which will be used for subsequent diffusion steps, and $x_{1}^{pre}$ is the predictive enhanced complex spectrogram in the first diffusion step. 
The reason for not using the predicted complex spectrogram directly is to maintain the feature distribution of the diffusion complex spectrograms. 
The two enhanced complex spectrograms are fused in the final step to exploit the complementary information: 
\begin{equation}
	\setlength{\abovedisplayskip}{0pt}
	\setlength{\belowdisplayskip}{0pt}
	\widehat{x_{n}} = \beta * x_{n} +  (1 - \beta) * x_{n}^{pre}
\end{equation}
where $	\widehat{x_{n}}$ is the final enhanced feature, and $x_{n}^{pre}$ is the predictive enhanced feature in the final diffusion step. 
The first and final fusion are only used for reverse diffusion process. 
The final loss combines both predictive and generative loss in (\ref{genloss}) and (\ref{preloss}).

\begin{figure}[h]
	\centering
	\begin{minipage}[t]{0.48\textwidth}
		\centering
		\includegraphics[width=1.\textwidth]{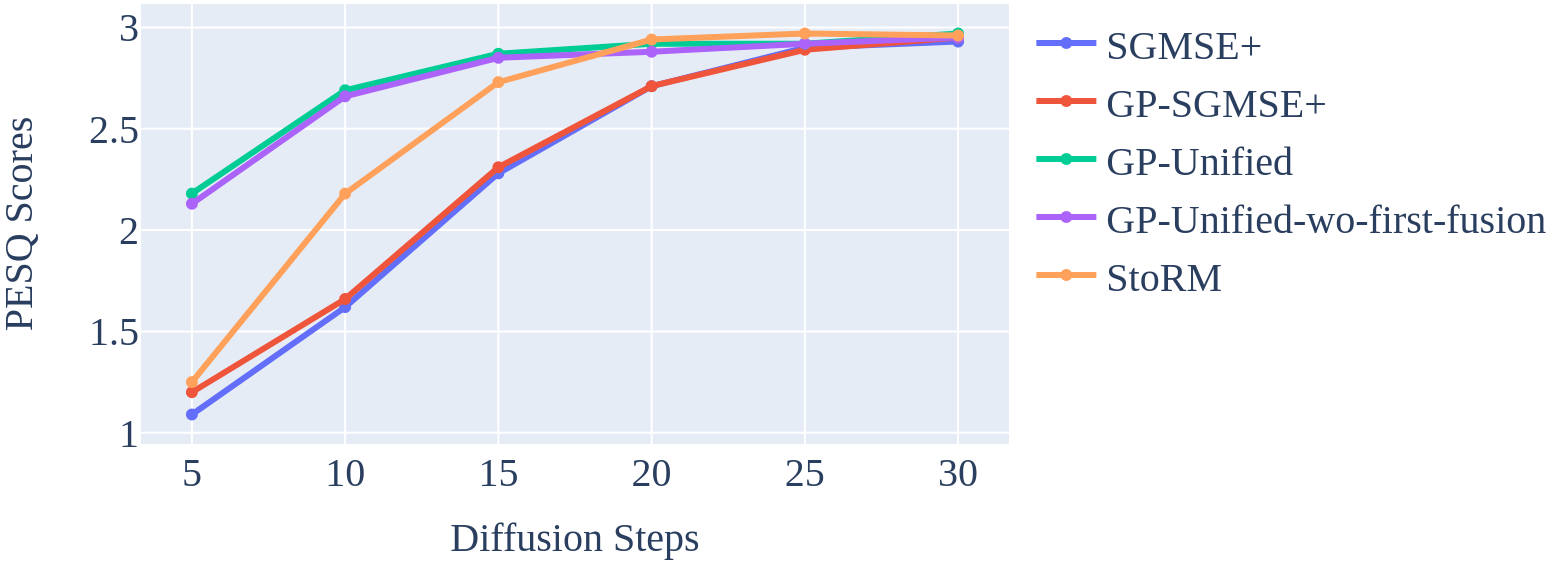}
		\vspace{-25pt}
		\caption{PESQ performance with different diffusion steps}
		\vspace{5pt}
		\label{pesq}
	\end{minipage}
	\begin{minipage}[t]{0.48\textwidth}
		\centering
		\includegraphics[width=1.\textwidth]{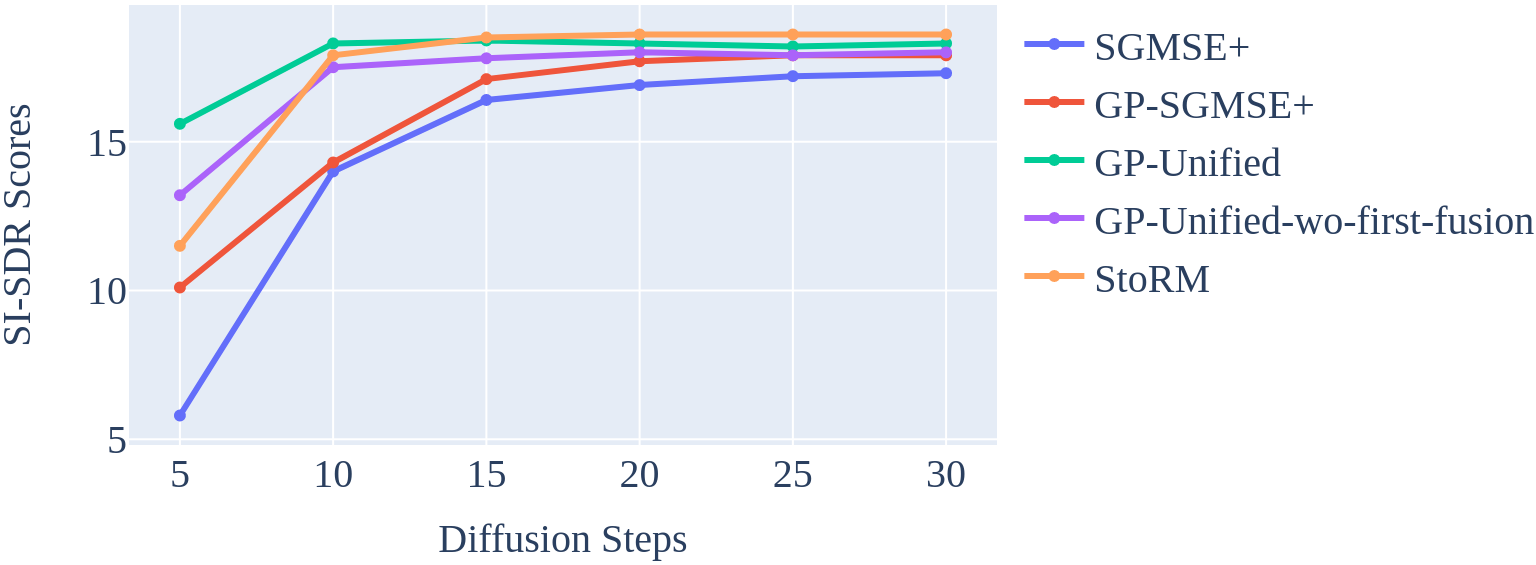}
		\vspace{-25pt}
		\caption{SI-SDR performance with different diffusion steps}
		\label{sisdr}
		\vspace{5pt}
	\end{minipage}
	\begin{minipage}[t]{0.48\textwidth}
		\centering
		\includegraphics[width=1.\textwidth]{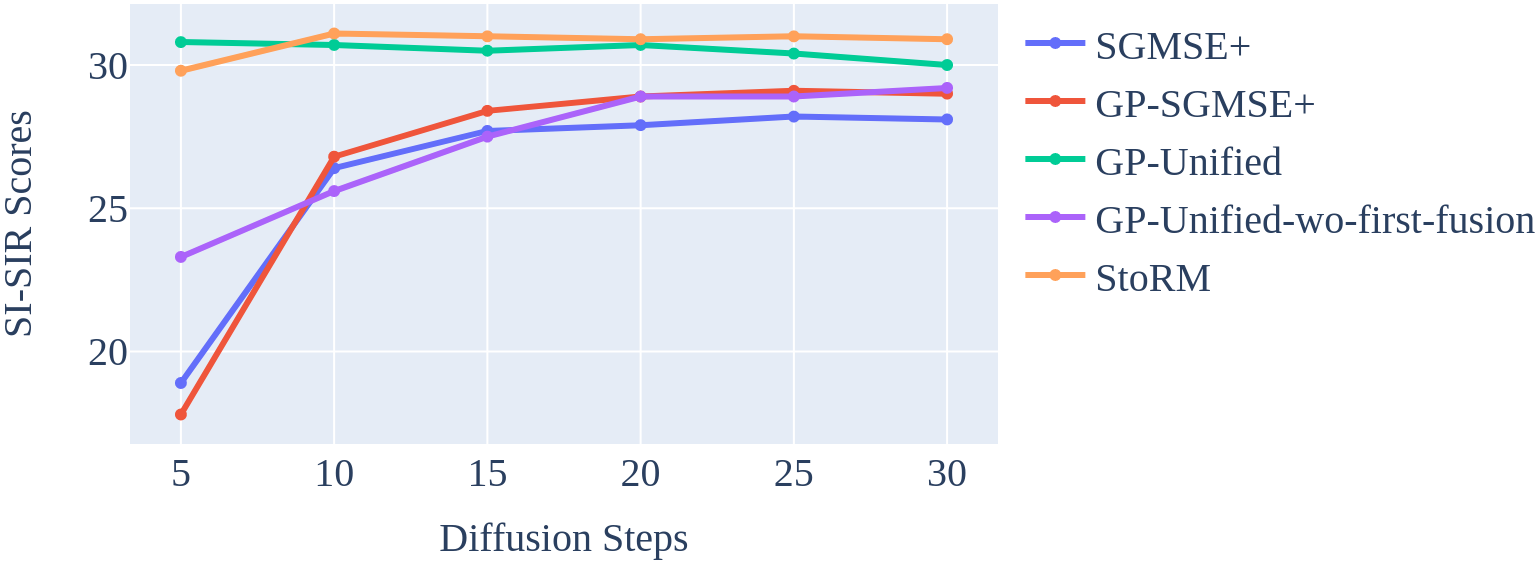}
		\vspace{-25pt}
		\caption{SI-SIR performance with different diffusion steps}
		\vspace{5pt}
		\label{sisir}
	\end{minipage}
	\begin{minipage}[t]{0.48\textwidth}
		\centering
		\includegraphics[width=1.\textwidth]{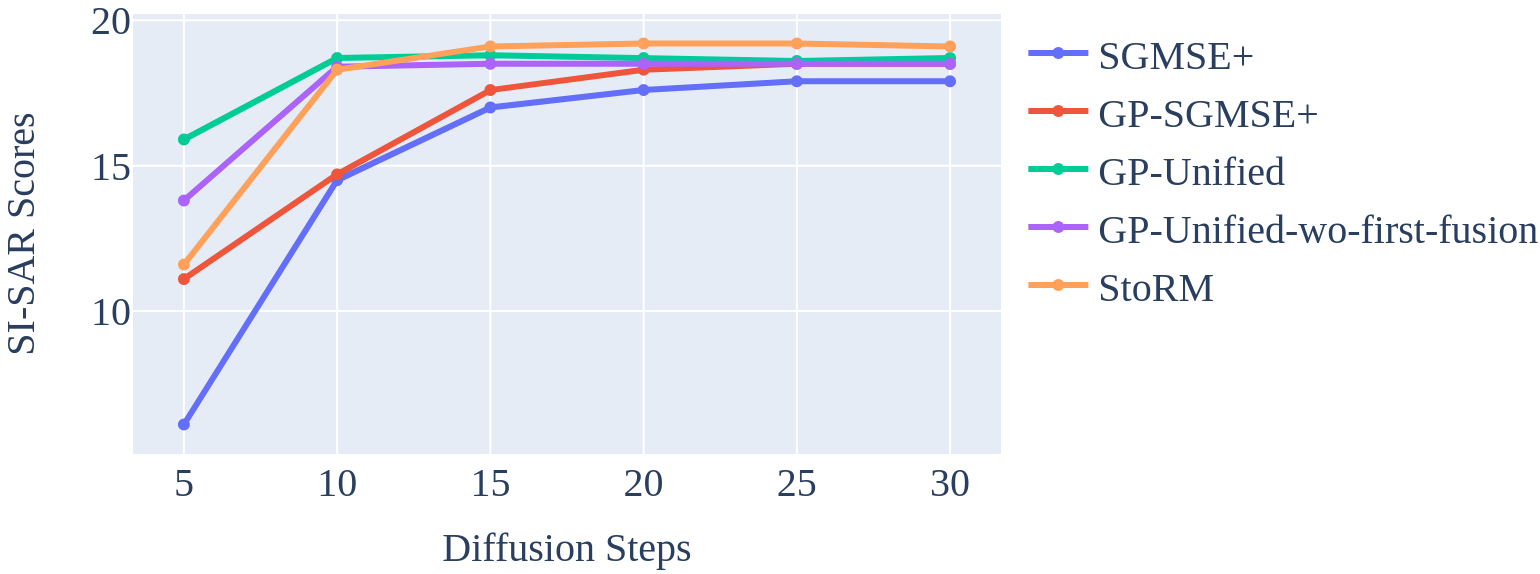}
		\vspace{-25pt}
		\caption{SI-SAR performance with different diffusion steps}
		\label{sisar}
	\end{minipage}
	\vspace{-20pt}
\end{figure}

\vspace{-5pt}

\section{Experimental Evaluations}
\vspace{-5pt}
The Noise Conditional Score Network (NCSN++\footnote{https://github.com/sp-uhh/sgmse}) architecture was used for the score model in both the baseline model (SGMSE+) and the proposed models (GP-SGMSE+, GP-Unified). 
The generative and predictive decoders had the same structure. 
The real and imaginary parts of the complex spectrograms were used as inputs. 
The residual blocks in upsampling and downsampling layers were based on the BigGAN architecture. 
Each upsampling layer consisted of three residual blocks, and each downsampling layer consisted of two blocks with the last block performing the upsampling or downsampling. 
Global attention was added at a resolution of $16 \times 16$ and in the bottleneck layer. 
All models were trained for 100 epochs. 

The experiments were based on the public Voicebank-DEMAND \cite{valentini2016investigating}. 
The dataset can be accessed from this {URL}\footnote{https://datashare.ed.ac.uk/handle/10283/1942}. 
All speech data were sampled at 16 kHz. 


To evaluate the proposed method, perceptual evaluation of speech quality (PESQ) \cite{rix2001perceptual}, extended short-time objective intelligibility (ESTOI) \cite{7539284}, scale-invariant signal-to-distortion ratio (SI-SDR) \cite{8683855}, scale-invariant signal-to-interference ratio (SI-SIR) \cite{8683855}, and scale-invariant signal-to-artifact ratio (SI-SAR) \cite{8683855} are used as the evaluation metric. 
Besides, the real-time factor (RTF) is used to evaluate the efficiency of different systems. 


We set the hyperparameter $\alpha$ for the first step fusion from 0.1 to 0.9, and finally found that 0.2 was the best. 
The hyperparameter $\beta$ for the final step fusion was 0.1. 
Because the two tasks are equally important, the weights of the training losses of the two parts are 0.5. 
We also tried fusing complex spectrograms at each step, but did not obtain improved performance.

\vspace{-5pt}


\vspace{-5pt}
\subsection{Effect of incorporating predictive loss function}
\vspace{-5pt}
As shown by comparison of ``SGMSE+'' and ``GP-SGMSE+'' in Figure~\ref{sisdr} to Figure~\ref{sisar}, 
introducing a predictive loss function into the diffusion model can effectively reduce speech distortion, reduce noise, and reduce artificial noise. 
However, the predictive information has a minor effect on the PESQ, as shown in Figure~\ref{pesq}.

\begin{table}[h]
	\centering
	\caption{PESQ performance of the predictive output of ``GP-Unified'' in different diffusion steps.}
	\begin{tabular}{lcccccc}
		\toprule
		Diffusion steps & 5 & 10 & 15 & 20 & 25 & 30  \\ 
		\midrule
		PESQ & 2.16 & 2.67 & 2.85 & 2.91 & 2.93 & 2.95   \\
		\bottomrule
	\end{tabular}
 \vspace{-5pt}
	\label{predpesq}
\end{table}

\begin{figure}[htbp]
	\centering
	\includegraphics[width=8.5cm]{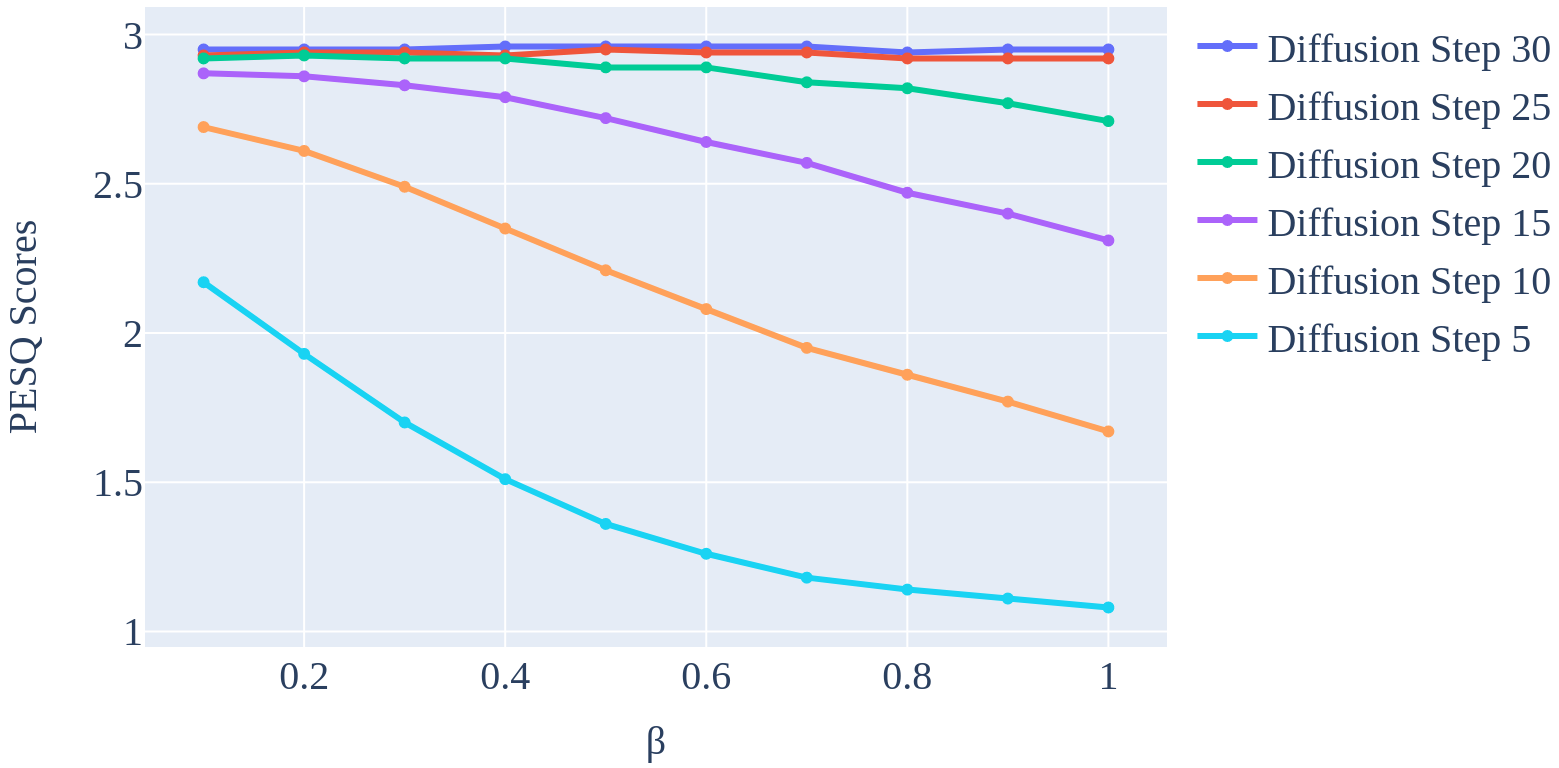}
	\vspace{-15pt}
	\caption{PESQ performance of ``GP-Unified'' under different diffusion steps with different $\beta$.}
	\label{beta}
	\vspace{-10pt}
\end{figure}

\vspace{-5pt}

\subsection{Effect of the first and final fusion}
\vspace{-5pt}
``GP-Unified-w/o-first-fusion'' adopts only the final fusion without the first diffusion-iteration fusion. 
As shown by the comparison of ``GP-Unified'' and ``GP-Unified-w/o-first-fusion'' in Fig.~\ref{sisdr} and \ref{sisar}, the first step fusion mainly affects the diffusion speed. 
Besides, Fig.~\ref{sisir} shows that the ``GP-Unified'' outperforms ``GP-Unified-w/o-first-fusion'' even though the model already has large diffusion steps. 
This implies that the first step fusion not only plays the role of initialization but also compensates for some speech distortion caused by diffusion.

``GP-SGMSE+'' and ``GP-Unified-w/o-first-fusion'' were trained by the same manner; the difference lies in whether they are fused in the final diffusion step. 
The final fusion step gives a significant PESQ improvement, especially when the number of diffusion steps is small, as shown in Fig.~\ref{pesq}. 
Furthermore, the final step fusion can significantly reduce speech distortion (Fig.~\ref{sisdr}) and artificial noise (Fig.~\ref{sisar})  when the number of steps is small. 
However, the effect is not obvious for noise suppression (Fig.~\ref{sisir}). 
Table~\ref{predpesq} shows the performance of the predictive output at different diffusion steps. 
The performance improves as the iteration steps increases. 
The proposed system can combine the characteristics of predictive and generative SE and fully use the predictive complex spectrograms even when the number of diffusion steps is small. 
When the diffusion steps increase, the complementarity between the generative and predicted  complex spectrograms is manifested. 
Through the fusion, PESQ and SI-SDR can be further improved, as shown in Fig.~\ref{pesq} and \ref{sisdr}.

\vspace{-5pt}

\subsection{Effect of fusion hyperparameter $\beta$}
\vspace{-5pt}
Figure~\ref{beta} shows the PESQ performance of ``GP-Unified'' under several diffusion steps with different values of $\beta$. 
$\beta$ has a more significant impact on smaller diffusion steps, and the advantages of the fusion are mainly reflected for smaller diffusion steps, especially in steps 5, 10, and 15. 
In step 5, the main performance improvement comes from predictive information. 
\begin{table}[htbp]
	\vspace{3pt}
	\centering
	\caption{Performance of different speech enhancement systems in VB dataset: ``Type'' denotes the type of the system,  ``P'' represents the predictive model, ``G'' represents the generative model.}
	\begin{tabular}{lcccc}
		\toprule
		System  & Type & PESQ & ESTOI & SI-SDR \\ 
		\midrule
		Mixture & -    & 1.97 & 0.79  & 8.4    \\ 
		\midrule
		Conv-Tasnet \cite{8707065} & P & 2.84 & 0.85 & 19.1 \\
		MetricGAN+ \cite{https://doi.org/10.48550/arxiv.2104.03538} & P & 3.13 & 0.83 & 8.5 \\  
		GaGNet \cite{LI2022108499} & P & 2.94 & 0.86 & \textbf{19.9}  \\
		\midrule
		SEGAN  \cite{Pascual2017} &  G    &   2.16   &  -     &    -    \\ 
		CDiffuSE \cite{9746901} & G & 2.46 & 0.79 & 12.6 \\ 
		SGMSE+ \cite{https://doi.org/10.48550/arxiv.2208.05830} & G &  2.93  &   0.87  &   17.3  \\ 
		StoRM \cite{https://doi.org/10.48550/arxiv.2212.11851}   & G &  2.93 & \textbf{ 0.88} & \textbf{18.8}  \\ 
		UNIVERSE\tablefootnote{Note that we implemented UNIVERSE ourselves because the code is not publicly available. The network was trained on VB, and we only considered mel bands for feature NLLs.} &G & 2.91 & 0.84 & 10.1 \\
		GP-SGMSE+ & G & 2.95 & 0.87 & 17.9 \\
		GP-Unified & G &  \textbf{2.97}  &  0.87  &  18.3   \\
		\bottomrule
	\end{tabular}
	\vspace{-10pt}
	\label{all_results}
\end{table}
In steps 10, 15, and 20, it mainly comes from the feature fusion: the performance of ``GP-Unified'' was improved compared to the generative (``GP-SGMSE+'') and the predictive model (results shown in the Table~\ref{all_results}). 
Their difference lies in the feature fusion. 
The system performs better when $\beta$ is smaller, which means that the generative information is incorporated into the predictive complex spectrogram.

\begin{table}[]
	\centering
	\caption{RTF with different evaluation metrics: ``\textbf{Score}'' represents the corresponding evaluation score.}
	\begin{tabular}{clccc}
		\toprule
		Evaluation Metrics      & Model      & Steps & Score & RTF  \\
		\midrule
		\multirow{3}{*}{PESQ}   & SGMSE+     & 30 & 2.93   & 1.68 \\
        & StoRM & 25 &2.93 & 1.4\\
		& GP-Unified & 15  & 2.93  & 1.35 \\
		\midrule
		\multirow{3}{*}{SI-SDR} & SGMSE+     & 30  & 17.3  & 1.68 \\
        & StoRM & 20 &18.6 &1.12 \\
		& GP-Unified & 10  &18.3   & 0.91 \\
		\bottomrule
	\end{tabular}
\vspace{-10pt}
\label{rft}
\end{table}

\vspace{-5pt}

\subsection{Comparasion with other methods}
\vspace{-5pt}
Comparison with other methods, including ``StoRM'' and ``UNIVERSE'', are listed in Table~\ref{all_results}. 
Compared with the proposed method, ``UNIVERSE'' significantly degraded SI-SDR. 
This suggests that incorporating predictive information by the form of ``UNIVERSE'' is not beneficial to improving segment-level performance. 
Compared with ``StoRM'', the proposed method had better PESQ performance. 
This suggests that the proposed method can achieve better frequency-domain enhancement performance, since PESQ depends on the frequency-domain performance. 
Table~\ref{rft} presents the real-time factor (RTF) values for "SGMSE+" and "GP-Unified". In comparison to "SGMSE+", "GP-Unified" achieves a substantial reduction in diffusion steps, resulting in improved RTF performance. Specifically, for PESQ, 15 diffusion steps are sufficient for convergence, saving 50\% of steps (RTF 1.35). For SI-SDR, 10 diffusion steps are adequate to achieve a comparable score of SGMSE+ and StoRM., resulting in a 66\% reduction in steps (RTF 0.91).
Compared with ``StoRM'', the RTF is still improved.

\section{Conclusions and Future Works}
In this paper, we have proposed a unified generative and predictive speech enhancement model (GP-Unified). 
The model encodes both generative and predictive information and applies the generative and predictive decoders separately, whose results are fused. 
The predictive information helps the model to reduce speech distortion, noise, and artifacts. 
The two systems are fused in the first and final steps. 
The information fusion can speed up the diffusion process by reducing the number of diffusion steps by about 50\%, which leads better RTF. 
Besides, information fusion can lead to better performance with the complementary between the predictive and generative SE.

\footnotesize
\bibliographystyle{IEEEtran}
\bibliography{refs}

\end{document}